\documentclass[10pt,letterpaper]{article}
\usepackage[top=0.85in,left=1.75in,footskip=0.75in,marginparwidth=2in]{geometry}

\usepackage[utf8]{inputenc}

\usepackage{cite}

\usepackage{nameref,hyperref}

\usepackage{microtype}
\DisableLigatures[f]{encoding = *, family = * }

\usepackage{changepage}

\usepackage[aboveskip=1pt,labelfont=bf,labelsep=period,singlelinecheck=off]{caption}

\makeatletter
\renewcommand{\@biblabel}[1]{\quad#1.}
\makeatother

\usepackage{lastpage,fancyhdr,graphicx}
\usepackage{epstopdf}
\fancyheadoffset[L]{2.25in}
\fancyfootoffset[L]{2.25in}

\usepackage{color}

\definecolor{Gray}{gray}{.25}

\usepackage{graphicx}

\usepackage{sidecap}

\usepackage{wrapfig}
\usepackage[pscoord]{eso-pic}
\usepackage[fulladjust]{marginnote}
\reversemarginpar

\begin{document}
\vspace*{0.35in}

\begin{flushleft}

\textit{This paper is an electronic preprint.}
\newline

{\Large
\textbf\newline{A discrete choice model for solving conflict situations between pedestrians and vehicles in  shared space}
}
\newline


Pascucci, F.\textsuperscript{1,*},
Rinke, N.\textsuperscript{2},
Schiermeyer, C.\textsuperscript{2},
Berkhahn, V.\textsuperscript{2},
Friedrich, B.\textsuperscript{1},

\bigskip
\bf{1} Technische Universit{\"a}t Braunschweig, 
Institute for Transportation and Urban Engineering, Hermann-Blenk-Stra{\ss}e 42, 38108 Braunschweig, Germany\\
\bf{2} Leibniz Universit{\"a}t Hannover, Institute for Risk and Reliability, Callinstra{\ss}e 34 30167 Hannover, Germany \\
\bigskip
* f.pascucci@tu-braunschweig.de

\end{flushleft}

\section*{Abstract}
When streets are designed according to the shared space principle, road user are encouraged to interact spontaneously with each other for negotiating the space. These interaction mechanisms do not follow clearly defined traffic rules but rather psychological and social principles related to aspects of safety, comfort and time pressure. However, these principles are hard to capture and to quantify, thus making it difficult to simulate the behavior of road users. 
This work investigates traffic conflict situations between pedestrians and motorized vehicles, with the main objective to formulate a discrete choice model for the identification of the proper conflict solving strategy.
A shared space street in Hamburg, Germany, with high pedestrian volumes is used as a case study for model formulation and calibration. Conflict situations are detected by an automatic procedure of trajectory prediction and comparison. Standard evasive actions are identified, both for pedestrians and vehicles, by observing behavioral patterns. A set of potential parameters, which may affect the choice of the evasive action, is formulated and tested for significance. These include geometrical aspects, like distance and speed of the conflicting users, as well as conflict-specific ones, like time to collision. A multinomial logit model is finally calibrated and validated on real situations. The developed approach is realistic and ready for implementation in motion models for shared space or any other less organized traffic environment. 


\section{Introduction}

The street design in the urban environment has to take account of the needs of a wide variety of road users, including many types of motor vehicles as well as vulnerable users like pedestrians and cyclists. When the street space has to accommodate them all together, traffic engineers may basically choose between two alternative design approaches. The first one is based on the separation of different transport modes. When traveling towards the same direction, separated traffic areas - e.g. separated with vertical kerbs - are established. When flows cross each other, control devices like markers, signs or signal devices are used to establish priority rules. The second one is based on the shared space principle and consists on designing a continuously paved surface with minimized road signs and markings. In these areas, road users are encouraged to interact spontaneously with each other for negotiating the space, hence taking priority or giving way to others and consequently adapting their trajectory and speed according to the traffic situation. \\

\subsection{Problem statement and previous research} 
Besides specific guidelines which can assist in the design process of shared space on the basis of technical recommendation and real examples \cite{DepartmentofTransport2011,Flow2012,FGSV2014}, traffic engineers at present cannot rely on realistic microsimulation tools, which would be useful for the Level of Service estimation and safety assessment. Despite that, in the recent years research has focused on shared space modeling by investigating the interaction mechanism among road users and proposing new methods to reproduce the behavior of people in this environments \cite{Schonauer2012,Anvari2015,Rinke2016}. The common approach has been to utilize the Social Force Model (SFM) of Helbing and Molnar \cite{Helbing1995} - originally developed for pedestrian dynamics - and to extend it to other road users like vehicles and cyclists. 
However, differently to pedestrian dynamic, the presence of motorized vehicle  makes all road users more vigilant due to the higher risk of injuries involved. Complex psychological processes are hiding behind each decision, e.g. whether to brake or not when a pedestrian is approaching on the side of the road. These very conflict situations among different transport modes are hard to reproduce and require further adaptation of the SFM - an approach that many researchers in the recent past have chosen.

Anvari et al. \cite{Anvari2014} extended the SFM with a conflict avoidance strategy, where the road user moving at the highest speed reacts first by decelerating or deviating and the other one reacts accordingly. That means, the model assigns priority to the weaker user regardless of the circumstances. Pascucci et al. \cite{Pascucci2015} used an algorithm to compare the future positions of the conflicting users, in order to assign priority to the one that would leave the conflict zone first. Sch{\"o}nauer et al. \cite{Schonauer2012} developed a tactical model which handles conflict situations by using a Stackelberg game, a rational game play which determines the winner by comparing single utilities of the players. The model considers parameters like the probability of collision, the distance between users and road regulation, with the purpose to determine the most probable reaction. However, as the authors declared, the estimated parameters were not presented due to the small amount of decisions collected on the field. 

Past research has also investigated pedestrian-vehicle conflicts in lane-based environments, i.e. where the behavior of road users is expected to be on a predefined trajectory \cite{Varhelyi1998,Oxley2005}. Many authors have proposed discrete choice models to determine the reaction performed by road users in pedestrian-vehicle conflicts \cite{Himanen1988,Sun2002,Schroeder2008}. In the field of shared space, Kaparias et al. \cite{Kaparias2016} investigated the pedestrian gap acceptance behavior by analyzing the effect of variables such as waiting time, crossing time, crossing speed and critical gap.



\subsection{Objective and contribution of this research} 

In the research project MODIS (Multi mODal Intersection Simulation) the authors have dealt with the issue of shared space microsimulation and have proposed a Social-Force based approach to simulate the interaction among road users in these areas \cite{Pascucci2015,Rinke2016,Schiermeyer2016}. Nevertheless, dealing particularly with pedestrian-vehicle conflicts revealed that road users base the decision whether an evasive action had to be performed on several criteria. To establish the impact of the respective criteria and to assess under which circumstances a specific evasive action is chosen, an extension of the proposed deterministic model would be necessary.

For this reason, this paper investigates the mechanism of reaction in pedestrian-vehicle conflicts in shared space, and aims to propose a discrete choice model to determine the most probable evasive action of a road user in given circumstances. The developed modeling approach has many innovative aspects in the field of shared space microsimulation: 

\begin{itemize}
	\item It provides decisions for single conflicting users. While in previous models a comprehensive conflict solving strategy is used, in the current approach every user involved in a conflict situation decides for themselves based on personal perception and intention. This can lead to situations where for example both users temporarily decide to give way each other for safety reasons (which is quite common in shared space).     
	\item It allows two possible reactions besides the possibility not to perform any reaction at all. Previous models were formulated on the binomial decision \textit{to accept/not to accept} the temporal gap between two consecutive cars, while here it is also specified how the evasive action is supposed to be. In this case, a multinomial logit is used to allow the user a total of three possible choices.  
	\item It considers multiple conflict situations. The number of simultaneous conflicts is indeed considered as an input variable in the proposed model.   
	\item The model is calibrated and validated on a large dataset of real world observations and is ready-to-use within a motion model for shared spaces or any other less-organized environment.  
\end{itemize}

The coefficients presented at the end of this paper are representative of the scenario where the model was calibrated and reflect the specific street layout, the road regulation and the local behavioral attitudes. Despite that, the developed procedure for coefficient estimation can also be used in different context from the one considered, including pedestrian crossings or other situations where priority is somehow negotiated.   

\subsection{Outline of the paper} 
This work is organized in four main steps, which reflect the methodology used to reach the objectives stated above: 
\begin{itemize}
\item a visual analysis is performed on real pedestrian-vehicle conflicts in order to identify, on a general level, which factors may affect road user's reaction choice;  
\item data are collected, including the determination of predictors and outcome for a set of selected conflict situations;
\item the significant variables are identified and the multinomial logit model is calibrated and validated, both for pedestrians and vehicles;
\item the model is tested on real situations to determine the performances.
\end{itemize}

This paper ends with some considerations about the developed model and some ideas for future research.

\section{Site location and observations}

\begin{figure}[h]
	\centering
	\includegraphics[width=6cm]{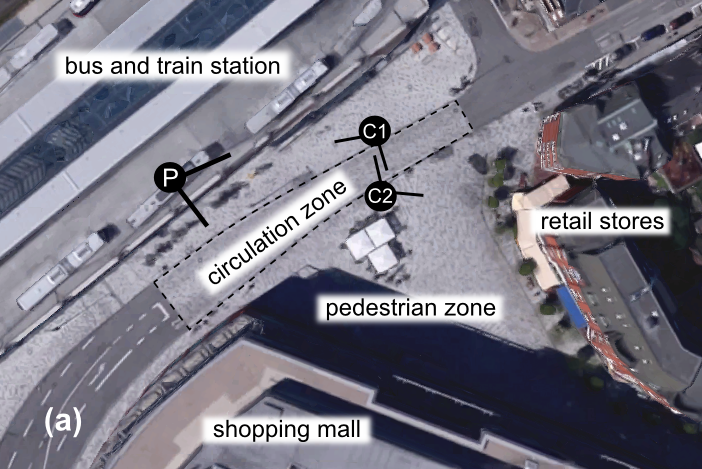}
	\includegraphics[width=6cm]{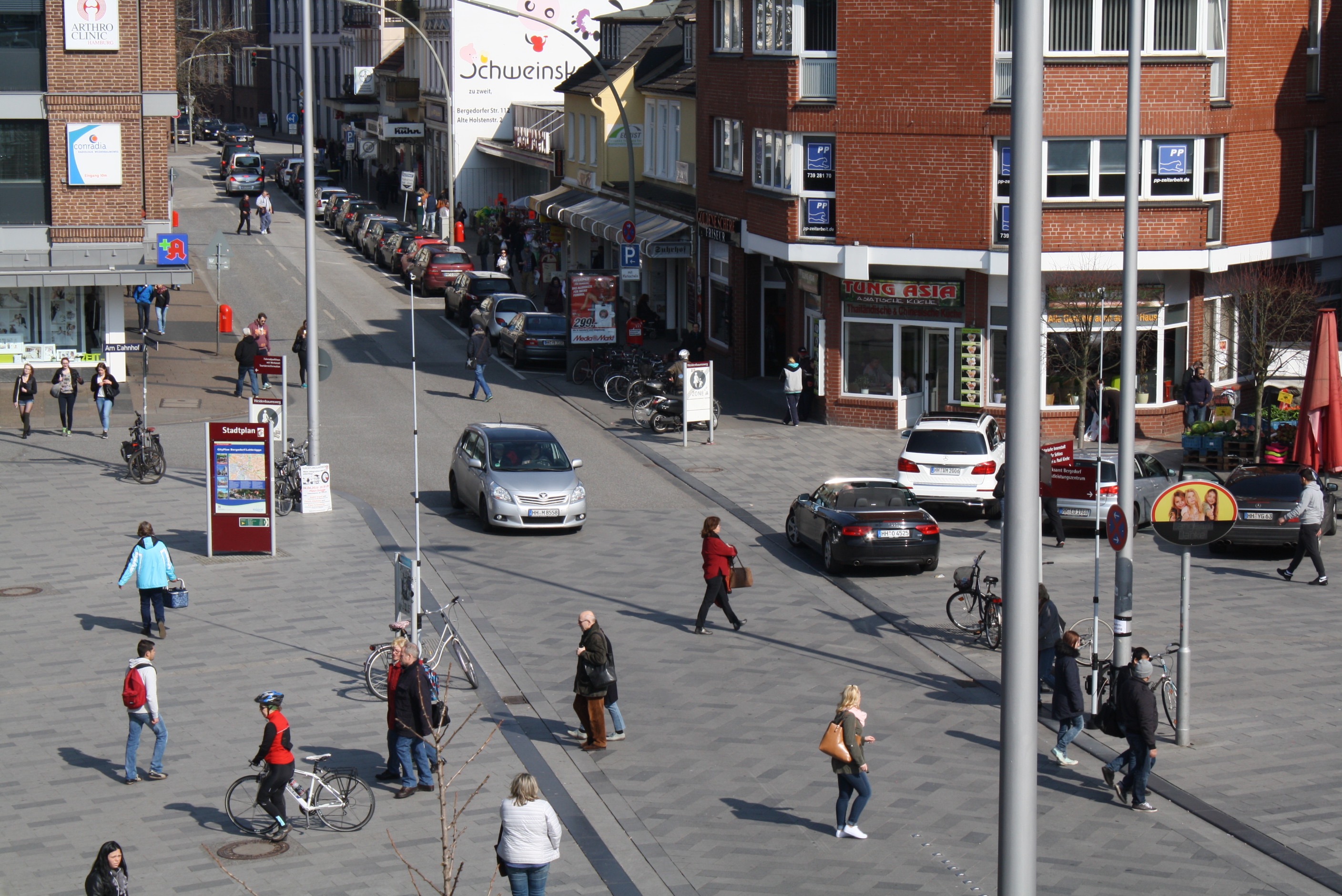}
	\caption{Site location. Aerial view (a) and picture from the bus station in point P (b)}
	\label{fig:eyebirdview}
\end{figure}

A shared space street in Hamburg (D) was chosen for video observation and data extraction. The site is located in proximity of the railway station of the district of Bergedorf (Weidebaumsweg) and has public space features due to the presence of retail stores, a shopping mall and a pedestrian zone (Fig.\ref{fig:eyebirdview}). This leads to high pedestrian crossing volumes over the 63-meter long area where vehicles are supposed to drive through (in this paper referred to as \textit{circulation zone}). To promote pedestrian movement all over the area, the shared space design principle has been adopted: a grey tone paved surface has been designed, which covers the \textit{circulation zone} and the surrounding pedestrian ones, with no-level difference between them but simply different patterns to identify the borders.

Within the \textit{circulation zone} vehicles are allowed to drive at maximum 20 km/h, while owning priority over crossing pedestrians coming from both sides, who have to wait for a sufficient gap between vehicles to cross the road. Despite that, the data survey has shown that vehicles and pedestrians negotiate the space spontaneously, often giving priority to each other as a courtesy rather than strictly follow road regulation. 

Video survey were conducted with two video cameras, placed at opponent borders of the circulation zone (points C1 and C2 of Fig.\ref{fig:eyebirdview}a), across from each other, at an elevation of about 7 meters. The video cameras have a resolution of 640 x 480 at 30 frames per second. The period of video recording was chosen as Saturday afternoon in the springtime (April 2nd 2016 from 2 to 5 pm), in order to maximize the volumes of crossing pedestrians. Besides vehicles and pedestrians, a few motorcycles and cyclist were present but their presence was negligible. 
To get an impression about the mechanisms of conflict reaction and the possible factors influencing them, typical pedestrian-vehicle conflict situations were extracted and visually analysed.

\subsection{Evasive action analysis}  
When entering an area shared with vulnerable road users, vehicle usually assume a different driving behavior, consisting of reducing speed and paying higher attention while driving. If no pedestrian is on either side a constant speed is kept until the end of the \textit{circulation zone}, but if pedestrians appear a decision is needed whether to give them way or not. In the first case, the typical reaction consists of simply decelerating, because weaving is usually considered to be dangerous. For vehicles, three different possible choices are assumed in this work: apart from the no reaction-behavior (No Reaction Vehicle, NRV), vehicles may decelerate - slightly or clearly - (Deceleration, DEC) or retrieve the desired speed by accelerating after a conflict situation (Acceleration, ACC).  

When no vehicle is present in the \textit{circulation zone}, pedestrians use to cross it more or less perpendicularly to the road axis in order to reduce the duration of the crossing within the \textit{circulation zone}. If a vehicle - or even more vehicles - appears, the basic question is whether to take priority or to give way. In both cases, pedestrian may choose not to react (No Reaction Pedestrian, NRP), because the conflict is assumed not to be dangerous or because a deceleration of the vehicle is expected. The second one is to react prudently (Prudent, PRU), by deviating parallel to the upcoming car and/or decelerating to keep a safe distance to the projected vehicle line. Moreover they can react aggressively (Aggressive, AGG) by deviating perpendicular to the trajectory of the vehicle, sometimes even by increasing the speed (in this way the pedestrian can leave the circulation zone earlier and allow the car to decelerate less intensively).  

\subsection{Possible relevant parameters}
Three main classes of parameters are assumed to be potentially influential on the choice of the evasive actions previously identified:
        
\begin{itemize}
	\item \textit{movement-specific}, like relative position, speed and acceleration of both road users;
	\item \textit{projected collision-specific}, which describe the expected situation if no evasive action is taken by any of the users;
	\item \textit{external conflict-specific}, related to the presence of other simultaneous situations of conflict;   
\end{itemize}
 
Many other parameters like age, gender and time pressure, which may possibly affect the behavior, are not considered in this work due to the difficulty to be captured in real world traffic situations.
In addition, no parameter describing the road layout or regulation is included, since only one scenario is considered. 

\section{Data acquisition and preparation}
In order to detect conflict situations automatically, the positions of all pedestrians and vehicles were tracked for a 30-min time interval at fixed time steps. To maximize the number of conflict situations, the most congested time interval was selected among the entire video material. The video tracking was manually performed by a time step of 0.5 seconds (15 video frames). The tracked points were transformed in the bird's eye view system and successively processed in three main phases.

The first phase was aimed at detecting conflict situations which are used for the model calibration. It consisted of a three-step methodology which was performed at every time step $ts^*$ of the time interval. Firstly, the expected behavior of road users was predicted by collecting the last 4 observed points of every road user, and by fitting a cubic smoothed spline, which can estimate the expected position in the next 8 seconds. Secondly, predicted positions were compared with each other to calculate the future relative minimum distance (MinDist) among road users. Thirdly, the current $ts^*$ - with the information of road user's ID - was saved as Conflict Instant (CI) when this distance was found to be below 5 meters. This resulted in a set of 2814 CIs, belonging to 409 conflict situations between one vehicle and one pedestrian. 

The second phase consisted of the computation of the predictors listed in Tab.\ref{tab:ExplanVar} at every time step.

\begin{table}[h]
	\caption{Set of explanatory variables considered}
	\label{tab:ExplanVar}
	\begin{center}
		\begin{tabular}{l l l l}
						Predictor & Type & Unit & Description \\\hline
			\textit{MinDist} & cont & m & minimum expected relative distance of road users\\
			\textit{TimeMinDist} & cont & t & temporal proximity to the situation of MinDist \\
			\textit{ActDist} & cont & m & distance at time step $ts^*$ between road users \\
		    \textit{OrtDist} & cont & m & distance at time step $ts^*$ between the pedestrian and \\& & &the expected 
			 trajectory of the vehicle \\
			\textit{TimeDelayXP} & cont  & s & temporal delay of the pedestrian - with respect to the  \\
			& & & vehicle - to reach XP (point where trajectories cross)** \\
			\textit{SpeedVeh} & cont  & m/s & speed of the vehicle at $ts^*$ \\
			\textit{AccVeh} & cont   & m/s\textsuperscript{2} & acceleration of the vehicle at $ts^*$  \\
			\textit{SpeedPed} & cont  & m/s & speed of the pedestrian at $ts^*$ \\
			\textit{AccPed} & cont  & m/s\textsuperscript{2} & acceleration of the pedestrian at $ts^*$ \\
			\textit{CPConfNr} & ord & & number of simultaneous conflict of a vehicle against \\& & & pedestrians\\
			\textit{PCConfNr} & ord & & number of simultaneous conflict of a pedestrian against \\& & & vehicles\\ 
			\textit{CarAhead} & bin & & if the driver has another vehicle behind\\ \hline
			\multicolumn{4}{ l }{\textit{**negative if the vehicle would anticipate the pedestrian  }} \\
		\end{tabular}
	\end{center}
\end{table}

\begin{figure}[h]
	\centering
	\includegraphics[width=6cm]{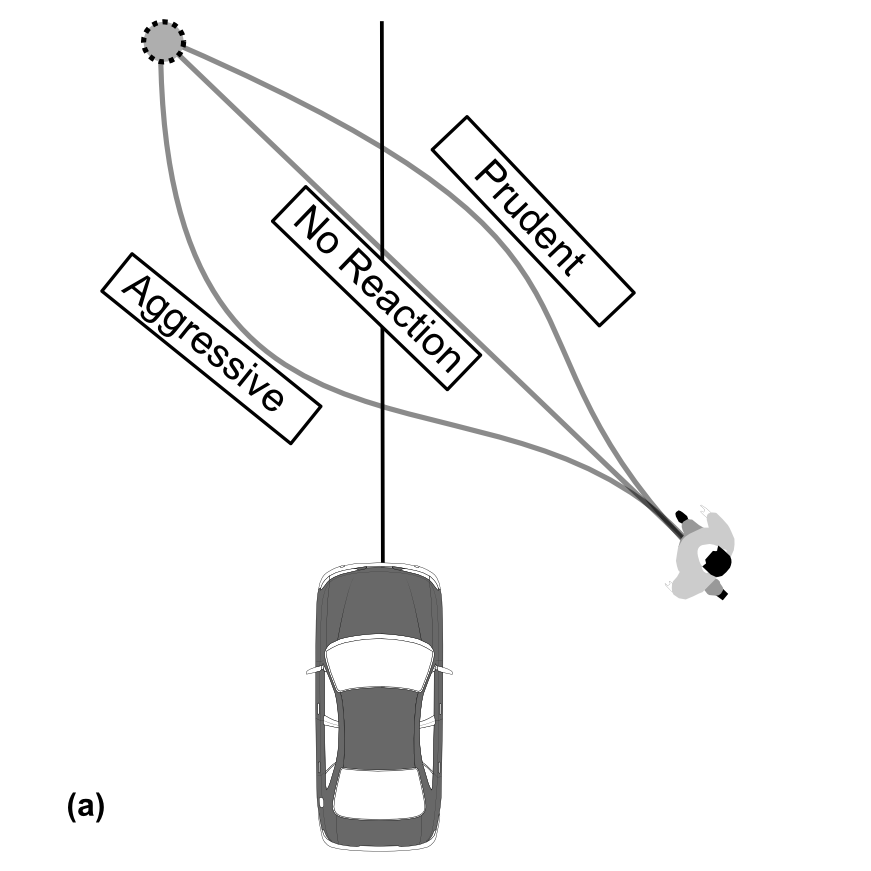}
	\includegraphics[width=6cm]{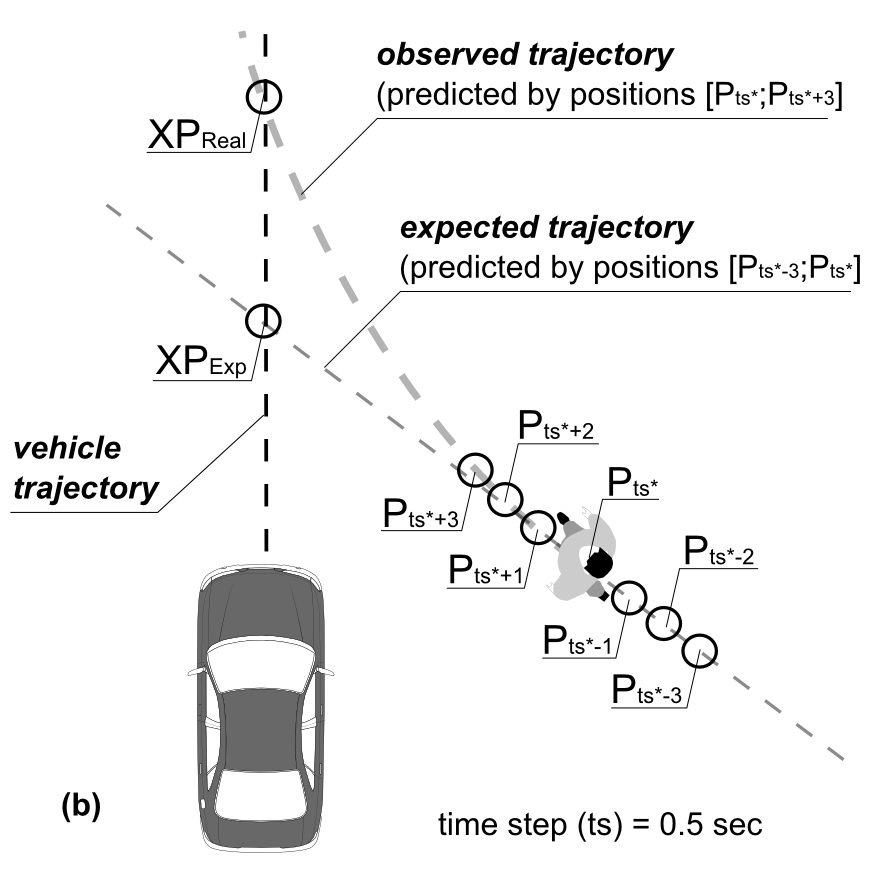}
	\caption{Reaction of a pedestrian in a conflict with a vehicle. Possible strategies (a) and detection of the reaction by the XP (b)}
	\label{fig:DecModExplanation}
\end{figure}

In the third phase it was determined how road users reacted to conflicts, i.e. which nominal outcome must be associated to every CI. For this purpose, the critical element is represented by the delay between the moment $ts^*$, where the conflict was observed, and the moment when the road user reacted accordingly. This temporal delay between stimulus and reaction is assumed here as 1.5 seconds, which includes the \textit{perception time} (needed to perceive the stimulus), the \textit{decision time} (needed to elaborate a conflict solving decision) and \textit{reaction time} (needed to react physically). Consequently, while the values of the predictors are calculated at time $ts^*$, the reaction choice is detected at time $ts^*+3 ts$ - i.e. 1.5 seconds later.
The type of evasive action is identified through a 5-step method which is briefly described here and shown in Fig.\ref{fig:DecModExplanation}b for the pedestrian case (and applies similarly to vehicles):

\begin{itemize}
	\item the \textit{expected trajectory} of the pedestrian is computed by a cubic smoothing spline through the last 3 observed positions and the actual one [$P_{ts^*-3}$ ; $P_{ts^*}$];
	\item the \textit{observed trajectory} of the pedestrian is computed by a cubic smoothing spline though the future 3 observed positions and the actual one [$P_{ts^*}$ ; $P_{ts^*+3}$];
	\item The intersections between the \textit{expected} and the \textit{observed trajectory} with the \textit{vehicle trajectory} are saved respectively as $XP_{Exp}$ and $XP_{Obs}$;
	\item The time needed by the pedestrian to reach $XP_{Exp}$ and $XP_{Obs}$ from $P_{ts^*}$ is computed;
	\item The temporal difference $k(ts^*)$ between $XP_{Exp}$ and $XP_{Obs}$ is used as the reference value to classify the reaction. Negative values of $k(ts^*)$ indicate that the pedestrian has adopted a prudent behavior with the intention to give way to the vehicle.  
\end{itemize}

\begin{figure}[h]
	\centering
	\includegraphics[width=6cm]{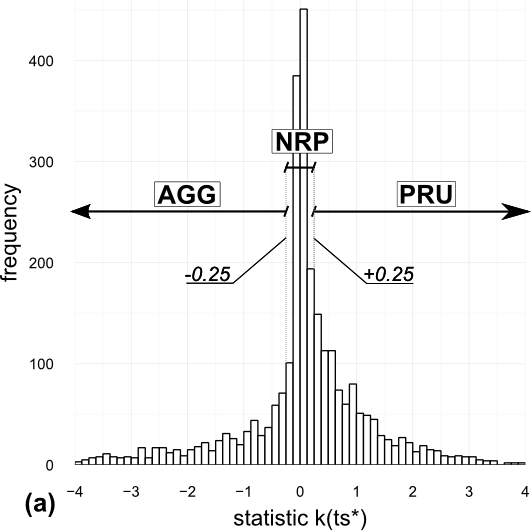}
	\includegraphics[width=6cm]{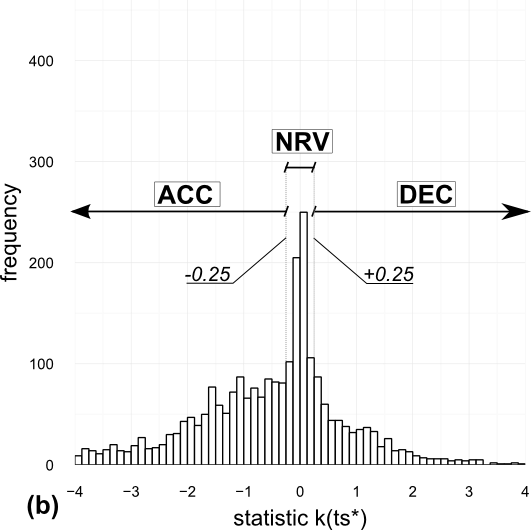}
	\caption{Distribution of statistic k for pedestrians (a) and vehicles (b) and assumed thresholds for determining the reaction}
	\label{fig:Histograms}
\end{figure}

The benefit of the statistic $k(ts^*)$ is due to the possibility to quantify the intensity of the evasive action by a single value, without computing any speed or directional change. The statistic $k(ts^*)$ was computed for all CIs, both for pedestrians and vehicles. The distribution of the variable is shown in the histograms in Fig.\ref{fig:Histograms}a for pedestrians and Fig.\ref{fig:Histograms}b for vehicles. Given the distribution of the variable $k(ts^*)$, arbitrary threshold values of $\pm 0.25 $ sec are assumed to determine which evasive action was chosen. In this way the reaction is classified and the dataset is ready for the model calibration.    

\section{Analysis and model fitting}

A multinomial logit model is chosen to investigate how each factor, among those previously identified, affects the choice of the evasive action of a pedestrian (or a motorist) when dealing with a conflict situation. This modeling approach was chosen for the categorical structure of the response, which may take three discrete values, and the type of predictors, which are both continuous and discrete. 

Taken the \textit{no reaction} as the baseline J - both for pedestrians and drivers - the multinomial logit assumes the log-odds of each alternative response $j$ to be linearly distributed with intercept $\alpha_k$ and a vector of regression coefficients $\beta_k$ (see Equation \ref{eq:LogReg}). The model allows the estimation of the probability $\mu_{CI\, j}$ in a certain CI to perform the evasive action $j$ given a set of explanatory variables $X_{CI}$.

\begin{equation} \label{eq:LogReg}
	ln(\mu_{CI\, j}/\mu_{CI\, J})= \alpha_k + X_{CI} \beta_k
\end{equation} 

Despite it is not excluded that some variables may have a quadratic or cubic relationship, a linear one was chosen with the purpose to keep the model as simple as possible.

\subsection{Selection of the predictors}
In a first stage, the dependence between the predictors and the outcome is tested, with the aim to identify a set of explanatory variables which are determinant for the choice of the evasive action. The analysis is carried out by maximum likelihood method and the statistical significance of each predictor is checked by the Z-value, which is defined as the regression coefficient divided by its standard error. The significance is checked with a two-sided test under the null hypothesis that the given variable does not affect the outcome. The analysis is performed both for pedestrian and drivers by estimating the coefficients $\beta_k$ for the model with all predictors (full model). This will highlight which variables are statistically significant and which could be omitted in the model. The results of the regression are shown in Tab. \ref{tab:VEHfullModel} for drivers and in Tab. \ref{tab:PEDfullModel} for pedestrians (in both cases one regression is performed for each alternative).

\begin{table}[tp]
	\caption{Vehicle decisional model, full model. Basemodel=No reaction }
	\label{tab:VEHfullModel}
	\begin{center}
		\begin{tabular}{l | l l l l | l l l l } 
	
	& \multicolumn{4}{ c }{No reaction to Prudent} & \multicolumn{4}{| c }{No reaction to Aggressive} \\
	
	Variable & $\beta_k$ & Std. err. & Z-value & Pr & $\beta$ & Std. err. & Z-value & Pr \\\hline
	\textit{MinDist}       & -0.38 & 0.06 & -6.48 & \textbf{0.00} 	& -0.28 & 0.07 & -4.22 & \textbf{0.00}	 \\
	\textit{TimeMinDist}   & 0.40 & 0.10 & 4.07 & \textbf{0.00}	    & 0.46 & 0.11 & 4.33 & \textbf{0.00}    \\
	\textit{ActDist}       & -0.02 & 0.03 & -0.85 & 0.40	    & 0.05 & 0.03 & 1.80 & 0.07    \\
	\textit{OrtDist}       & 0.16 & 0.06 & 2.55 & \textbf{0.01}       & 0.16 & 0.07 & 2.28 & \textbf{0.02}    \\
	\textit{TimeDelayXP}   & 0.17 & 0.02 & 8.36 & \textbf{0.00}       & 0.10 & 0.02 & 4.20 & \textbf{0.00}    \\	
	\textit{SpeedVeh}      & -0.05 & 0.09 & -0.59 & 0.55      & -1.03 & 0.12 & -8.50 & \textbf{0.00}    \\
	\textit{AccVeh}        & -1.70 & 0.11 & -14.88 & \textbf{0.00}      & 1.27 & 0.13 & 10.07 & \textbf{0.00}    \\
	\textit{SpeedPed}       & -0.18  & 0.23 & -0.79 &  0.43    & 0.24 & 0.27 & 0.90 & 0.37    \\
	\textit{AccPed}         & 0.69 & 0.28 & 2.44 &  \textbf{0.01}      & -0.92 & 0.34 & -2.72 &  \textbf{0.01}   \\
	\textit{CPConfNr}      & 0.14 & 0.05 & 2.84 & \textbf{0.00}       & 0.05 & 0.06 & 0.89 & 0.38    \\
	\textit{CarAhead}      & 0.58 & 0.39 & 1.50 & 0.13       & 0.20 & 0.48 & 0.42 & 0.67    \\	\hline
	\multicolumn{9}{ l }{\textit{Number of observations = 2,838; Dev = 3576.22; Constant-only model: Dev. = 5545.31 }} \\
	\multicolumn{9}{ l }{\textit{Goodness of Fit: $chi^{2}$= 1969.09 with d.f.=22. Prob$\ge$ $chi^{2}$ = 1 }} 
		\end{tabular}
	\end{center}
\end{table}

\begin{table}[tp]
	\caption{Pedestrian decisional model, all variables. Basemodel=No reaction}
	\label{tab:PEDfullModel}
	\begin{center}
		\begin{tabular}{l | l l l l | l l l l } 
			
			& \multicolumn{4}{ c }{No reaction vs Prudent} & \multicolumn{4}{| c }{No reaction vs Aggressive} \\
			
			Variable & $\beta_k$ & Std. err. & Z-value & Pr & $\beta$ & Std. err. & Z-value & Pr \\\hline
			\textit{MinDist}   & -0.53 & 0.06 & -8.81 & \textbf{0.00} 			& -0.32 & 0.05 & 6.04 & \textbf{0.00} 	 \\
			\textit{TimeMinDist}    & 0.54  & 0.11 & 5.80  & \textbf{0.00} 		& 0.62  & 0.08 & 7.25 & \textbf{0.00}	\\
			\textit{ActDist}    & 0.00 & 0.09 & 0.02  & 0.98	 					& -0.04 & 0.02 & -1.84 & 0.07\\
			\textit{OrtDist}  & 0.30 & 0.07 & 4.55  & \textbf{0.00} 				& 0.07 & 0.06 & 1.16 & 0.25\\
			\textit{TimeDelayXP}  & 0.21 & 0.02 & 8.64 & \textbf{0.00}		    & 0.23 & 0.02 & 10.02 & \textbf{0.00} \\
			\textit{SpeedVeh}  & -0-06 & 0.9 & -0.65 & 0.52 						& 0.21  & 0.08 & 2.66 & \textbf{0.01}  \\
			\textit{AccVeh}  & 0.32  & 0.09 & 3.53 & \textbf{0.00}  						& 0.12 & 0.08 & 1.46 & 0.15\\	
			\textit{SpeedPed}  & -0.53 & 0.26 & -2.05 & \textbf{0.04} 					& -2.41 & 0.23 & -10.41 & \textbf{0.00} \\
			\textit{AccPed}  & -4.01 & 0.36 & -11.13  & \textbf{0.00} 			& 2.48 & 0.30 & 8.39 & \textbf{0.00}  \\
			\textit{CPConfNr}  & -0.08 & 0.05 & -1.52 & 0.13						& -0.01 & 0.05 & -0.30 & 0.77 \\
			\textit{CarAhead}  & -0.30 & 0.35 & -0.87 & 0.39 					& 0.08 & 0.29 & 0.28 & 0.78 \\	
			\textit{PCConfNr}  &  -0.10 & 0.10 & -1.00 & 0.32						& -0.13  & 0.09 & -1.41 & 0.16 \\ \hline
			\multicolumn{9}{ l }{\textit{Number of observations = 2,838; Dev = 4035.30; Constant-only model: Dev. = 6133.63 }} \\
			\multicolumn{9}{ l }{\textit{Goodness of Fit: $chi^{2}$= 2098.33 with d.f.=22. Prob$\ge$ $chi^{2}$ = 1 }} 
		\end{tabular}
	\end{center}
\end{table}

The result of the goodness-of-fit test expressed through the chi-squared statistic shows that the improvement given by the explanatory variables with respect to the null model is significant. The associated p-value (calculated under the null hypothesis that the model fits the data well) is approximately 1 and this suggests that further model specifications - e.g. quadratic relations - are not necessary at the moment.
Looking at the p-values of single predictors, it can be noticed that only part of them are statistically significant independently from the road user and the chosen reaction, i.e. \textit{MinDist}, \textit{TimeMinDist}, \textit{TimeDelayXP} and \textit{AccPed} (p-values are close to 0). The sign also indicates clear tendencies: both road users tend to take an evasive action when \textit{MinDist} decreases, \textit{TimeMinDist} increases and \textit{TimeDelayXP} increases. That means that road users are more inclined to change their behavior when the moment of minimum closeness is temporarily distant, when it will imply a collision, and when they are ahead of the conflicting user. Moreover, they tend to behave more prudently when they are in a deceleration phase, and aggressive when they are accelerating. Other tendencies are user- and reaction-specific, like for example \textit{CPConfNr}, which is relevant only for drivers when deciding if to decelerate or not.
For the sake of the model's simplicity, part of the predictors are excluded. The selection was done by excluding variables once a time, and checking for consistent decreases in residual deviance. Relative chi square are respectively 1943.8 and 2052.7, which is very close to the one of the full model. 

\subsection{Model calibration and validation}
In order to calibrate and validate the model on different data, the entire sample was splitted into 70\% for training and 30\% for testing. The coefficient were estimated on the training sample (Tab.\ref{tab:Calibr}) and successively tested on the validation one, where the likelihood of every reaction choice was computed for all the CIs and the option with highest probabilities was assumed as the response. The results are shown in the confusion matrix (Tab.\ref{tab:ConfusionMatrix}), where each column represents the instances in the predicted class, while each row represents the instances in the observed class.

\begin{table}[h]
	\caption{Coefficients $\beta$ estimated by cross-validation}
	\label{tab:Calibr}
	\begin{center}
		\begin{tabular}{l l l|l l l} 
			\multicolumn{3}{ c }{Vehicle Model} & \multicolumn{3}{ c }{Pedestrian Model} \\
			Variable & NR to PR & NR to AG & Variable & NR to PR & NR to AG \\\hline
			\textit{Intercept}   & 0.196 & -0.309 & \textit{Intercept} & -2.193 & 1.057 \\
			\textit{MinDist}   & -0.402  & -0.265 & \textit{MinDist} & -0.497 & -0.309 \\
			\textit{TimeMinDist}  & 0.365  & 0.539 & \textit{TimeMinDist} & 0.745 & 0.547 \\
			\textit{OrtDist} & 0.136  & 0.225 & \textit{TimeDelayXP} & 0.288 & 0.252 \\
			\textit{TimeDelayXP}  & 0.161  & 0.116 & \textit{SpeedPed} & 0.099 & -2.344 \\
			\textit{SpeedVeh} & -0.118  & -0.800 & \textit{AccPed} & -3.919 & 2.484 \\
			\textit{AccVeh} & -1.738  & 1.199 & \textit{AccVeh} & 0.327 & 0.131 \\ 
			\textit{AccPed} & 0.659  & -0.882 & \multicolumn{3}{ c }{ } \\
			\hline			
		\end{tabular}
	\end{center}
\end{table}

\begin{table}[h]
	\caption{Confusion matrix, observed (rows) and predicted (columns) outcome}
	\label{tab:ConfusionMatrix}
	\begin{center}
		\begin{tabular}{l l l l|l l l l} 
		\multicolumn{4}{ c }{Vehicle Model} & \multicolumn{4}{ c }{Pedestrian Model} \\
				& NR  & Dec  & Acc      & 	   & NR  & Prud & Agg \\
			NR  & \textbf{125}  & 57  & 14    	& NR   & \textbf{262}  & 27	& 49	 \\
			Dec & 22  & \textbf{420} & 22		& Prud & 24  & \textbf{117}	& 71	 \\    
			Acc & 21  &	55	 & \textbf{92}		& Agg  & 54  & 39	& \textbf{199}	 \\   \hline
					
		\end{tabular}
	\end{center}
\end{table}

The off-diagonal elements of the confusion matrix reveal in which situations the predicted choice differs from the observed one. The misclassification rate, i.e. the percentage of off-diagonal elements with respect to the total, amounts to 23.1\% for drivers and 31.3\% for pedestrians. This has to be considered a satisfying result given the high stochasticity of road user's behavior, which may be strongly affected by parameters like age, sex or time pressure. 

\begin{figure}[h]
	\centering
	\includegraphics[width=4cm,height=4cm]{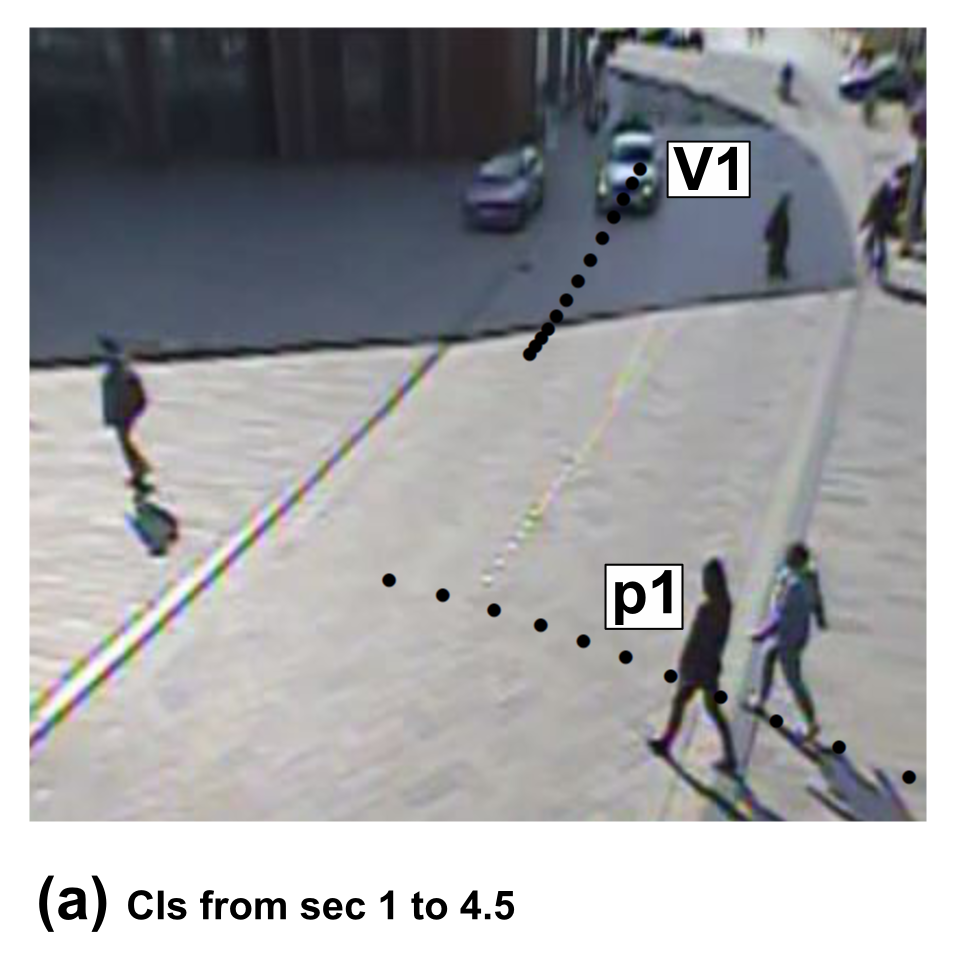} 	 
	\includegraphics[width=4cm,height=4cm]{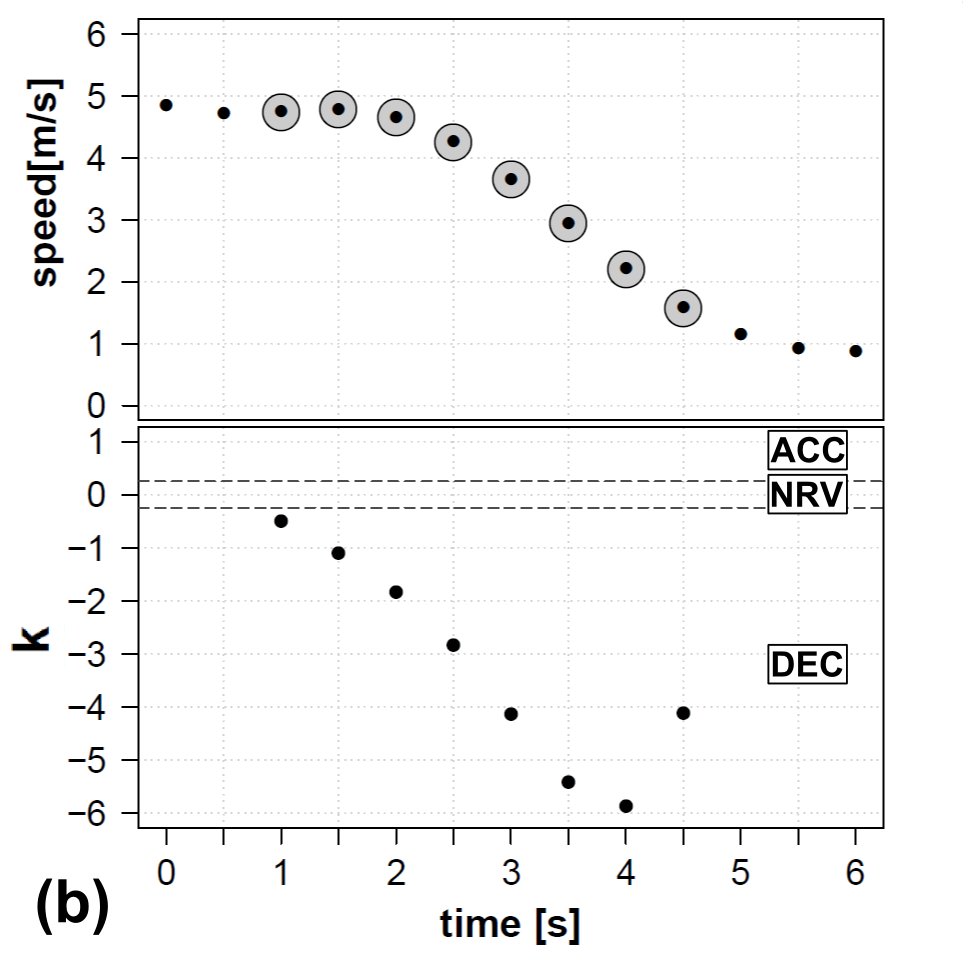}
	\includegraphics[width=4cm,height=4cm]{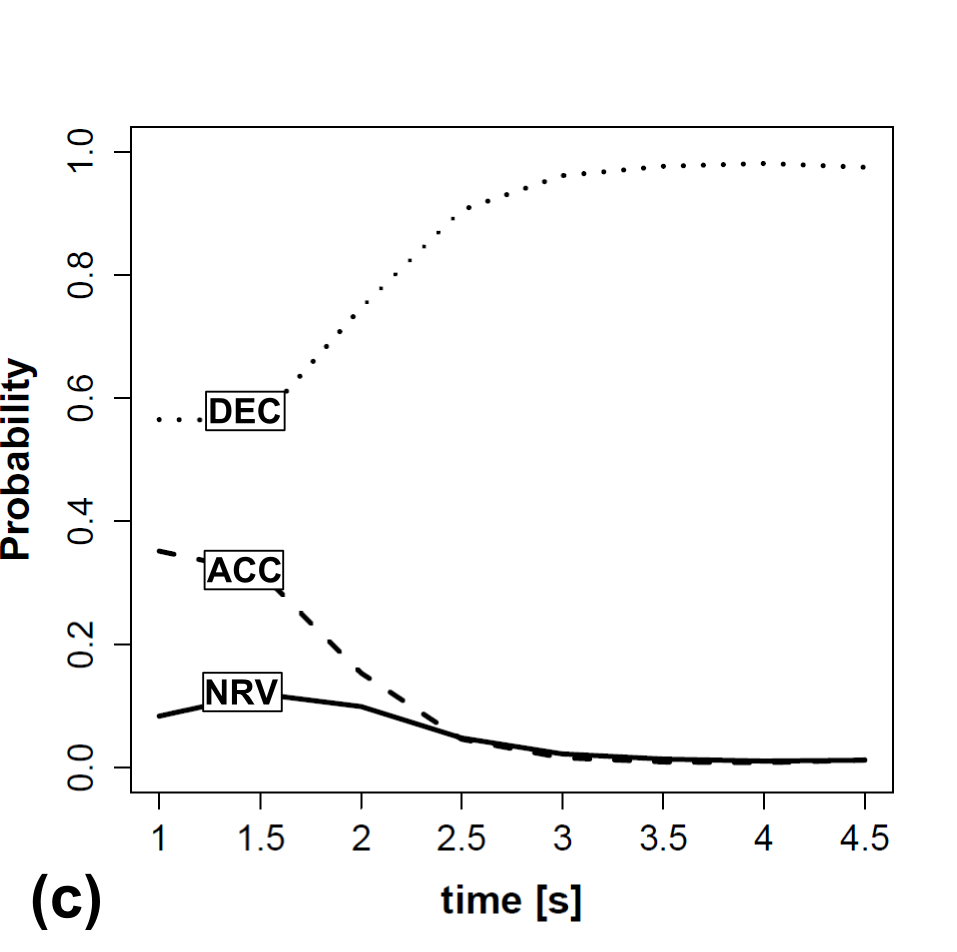}
	\caption{Situation 1: Characteristic values of vehicle V1. The model reproduces well the observed behavior.}
	\label{fig:Sit1}
\end{figure}

\begin{figure}[h]
	\centering
	\includegraphics[width=4cm,height=4cm]{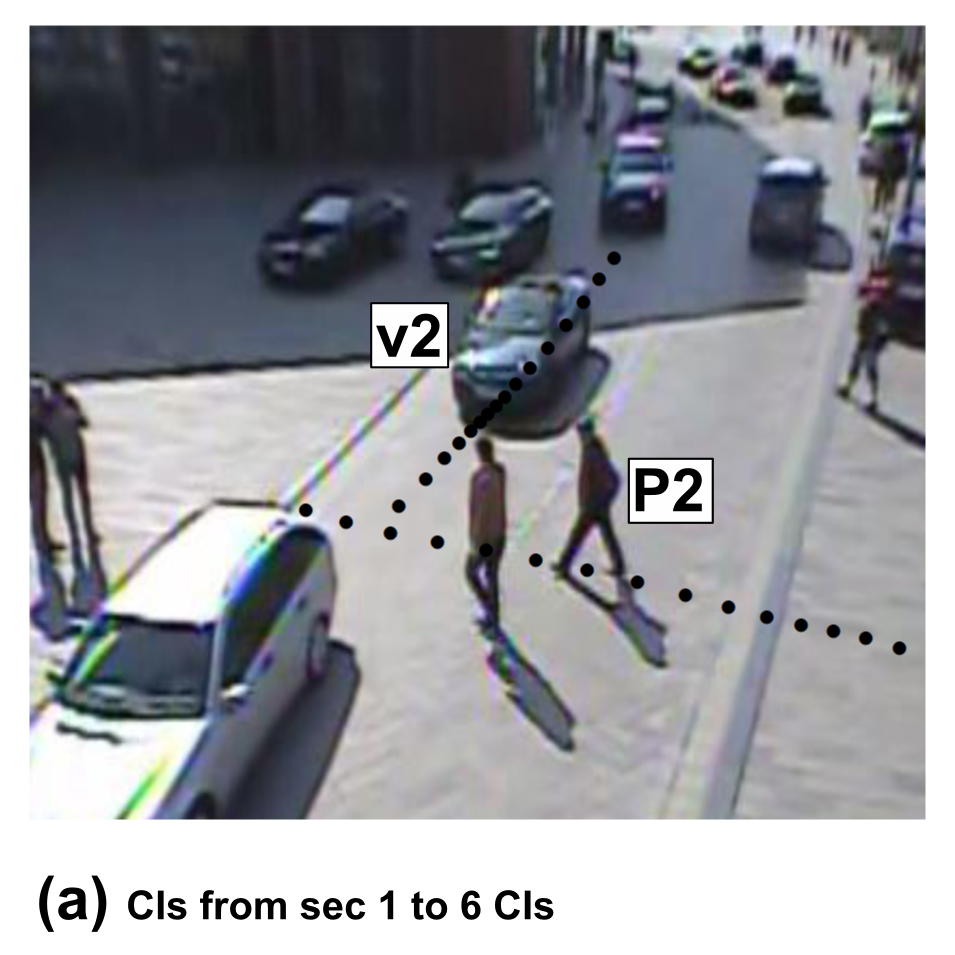} 	 
	\includegraphics[width=4cm,height=4cm]{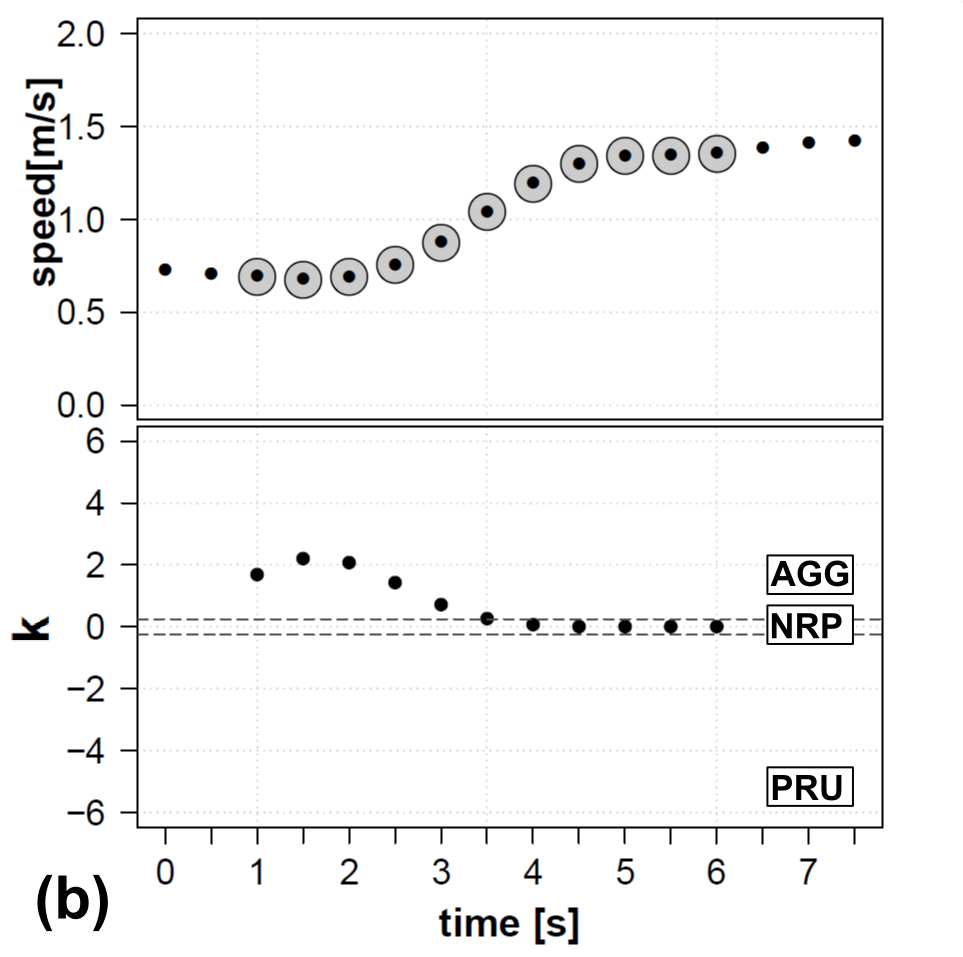}
	\includegraphics[width=4cm,height=4cm]{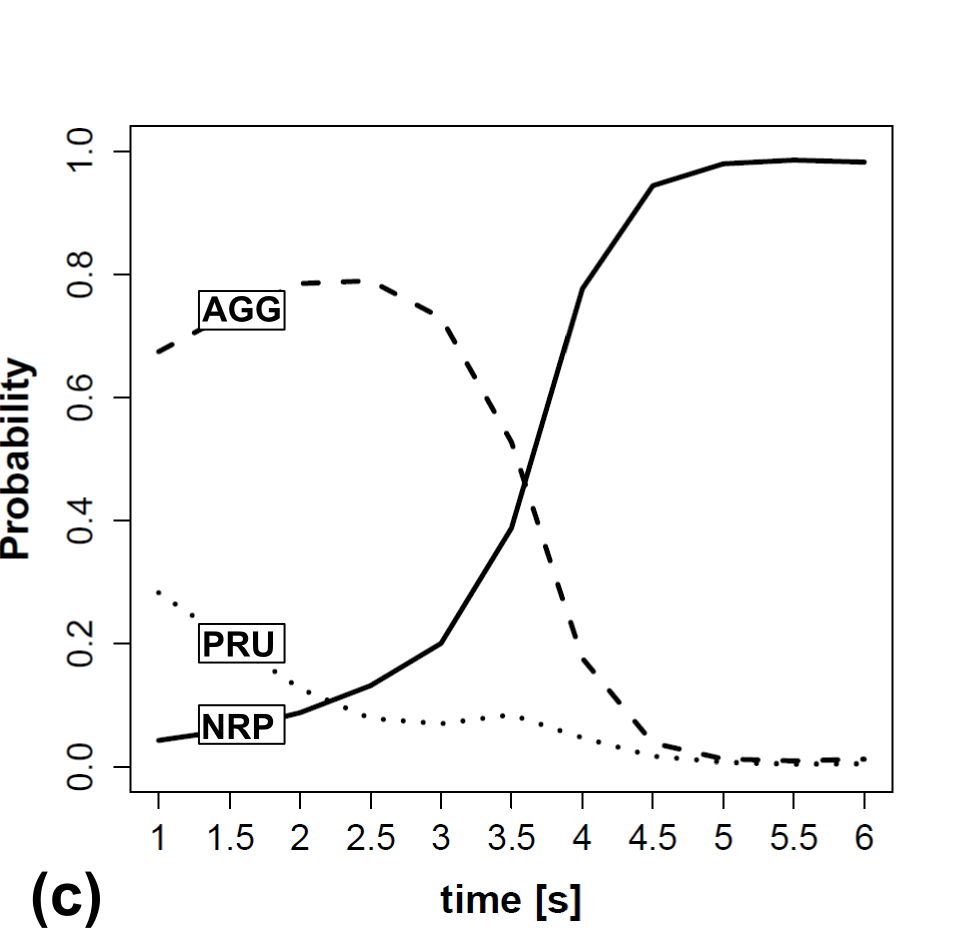}
	\caption{Situation 2: Characteristic values of pedestrian P2. The model reproduces well the observed behavior}
	\label{fig:Sit2}
	
\end{figure}

Two exemplifying situations are chosen to show the good performances of the developed model. For each situation three figures are shown alongside each other: a frame of the video sequence which displays the conflict dynamic (a), the observed behavior in terms of speed (or direction) with the associated value of the statistic $k$ for every CI (b) and the probabilities predicted by the model related to the different reaction choices (c). For the sake of clarity, pedestrian and vehicles are indicated by the letter $v$ and $p$. Moreover, these abbreviations are capital when the behavior of the road user is estimated and tested against observations. In situation 1 (Fig.\ref{fig:Sit1}) the vehicle $V1$ is in conflict with pedestrian $p1$ (as well the pedestrian next to $p1$). The latter decides to cross the \textit{circulation zone} and forces $V1$ to give way by decelerating. The model predicts the choice correctly for the whole time of conflict, since the DEC probability is always higher than the alternative ones. In situation 2 (Fig.\ref{fig:Sit2}) pedestrian $P2$ (as well as the one next to $P2$) step onto the roadway with a speed of approximately 0.75 m/s (lower as the desired speed). As the vehicle $v2$ decelerates to give him way - and also because there is a vehicle ahead - $P2$ accelerates until it reaches its desired speed (around 1.45 m/s). This behavior is identified as AGG by the statistic $k$ for the first part of the conflict, and is consistent with the outcome of the developed model. Moreover, the transition from AGG to NRP is reproduced very well.

\begin{figure}[h]
	\centering
	\includegraphics[width=4cm,height=4cm]{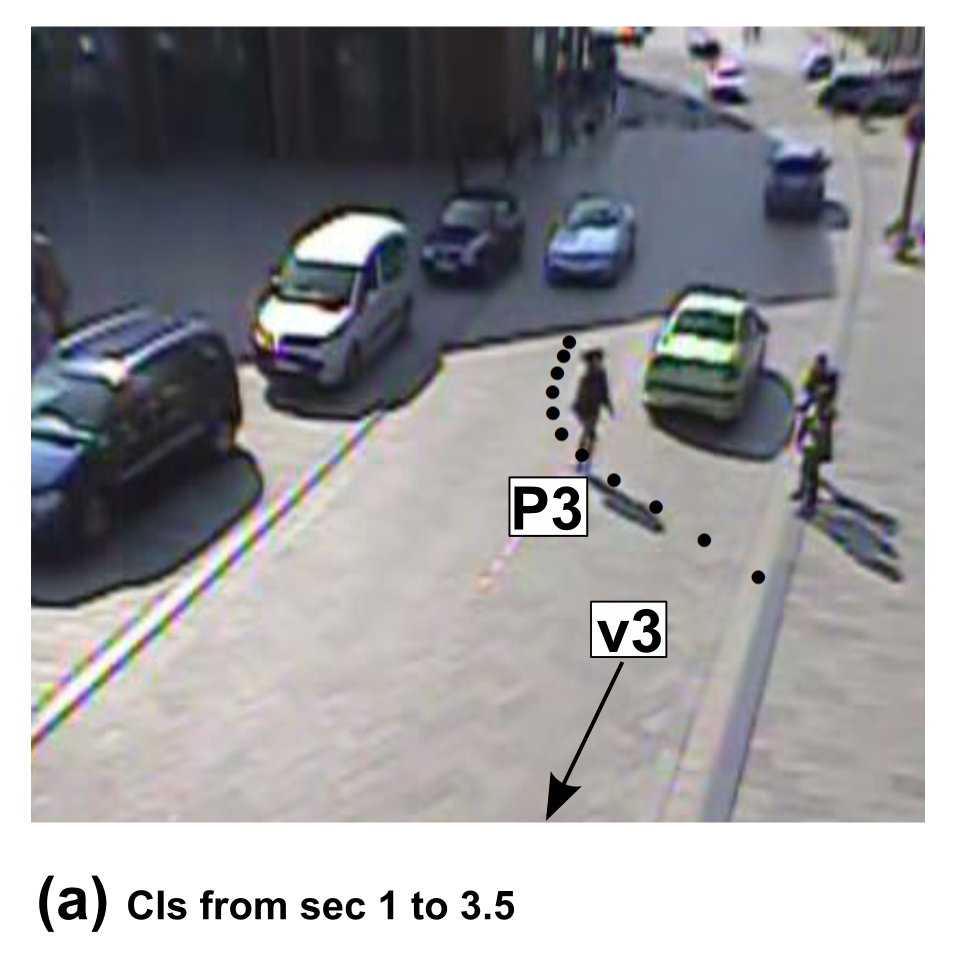} 	 
	\includegraphics[width=4cm,height=4cm]{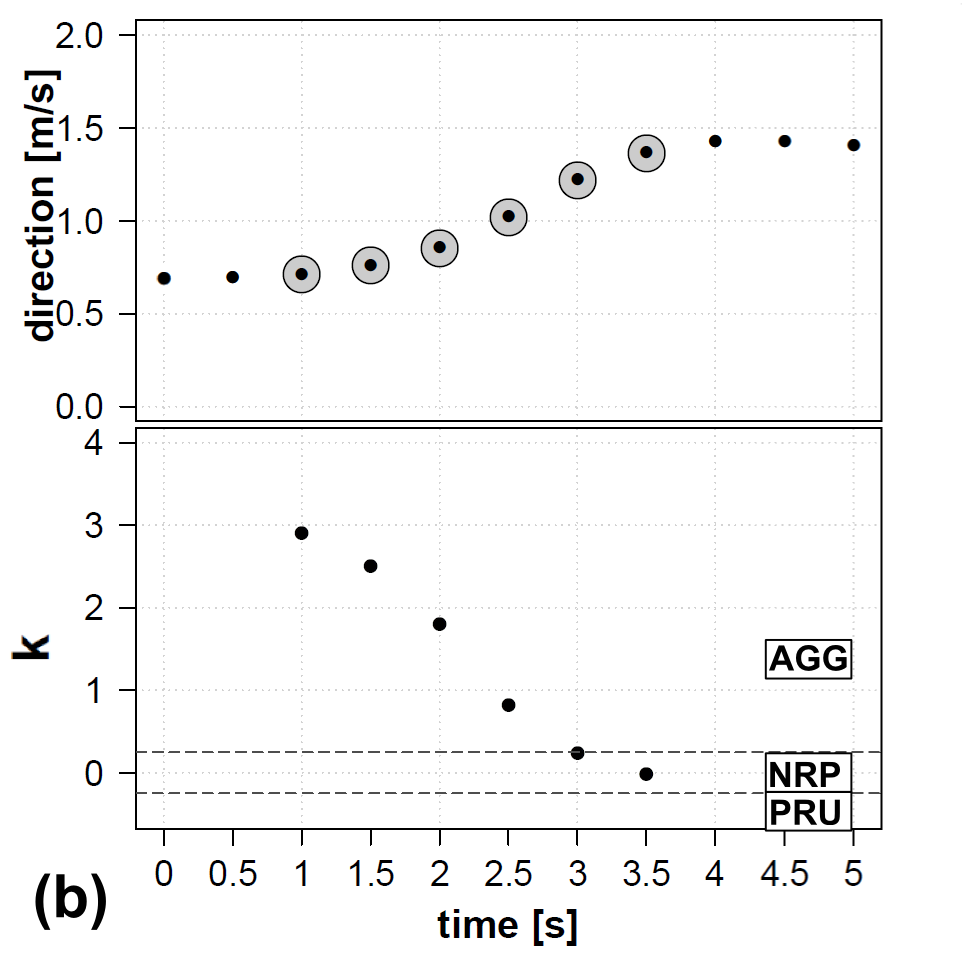}
	\includegraphics[width=4cm,height=4cm]{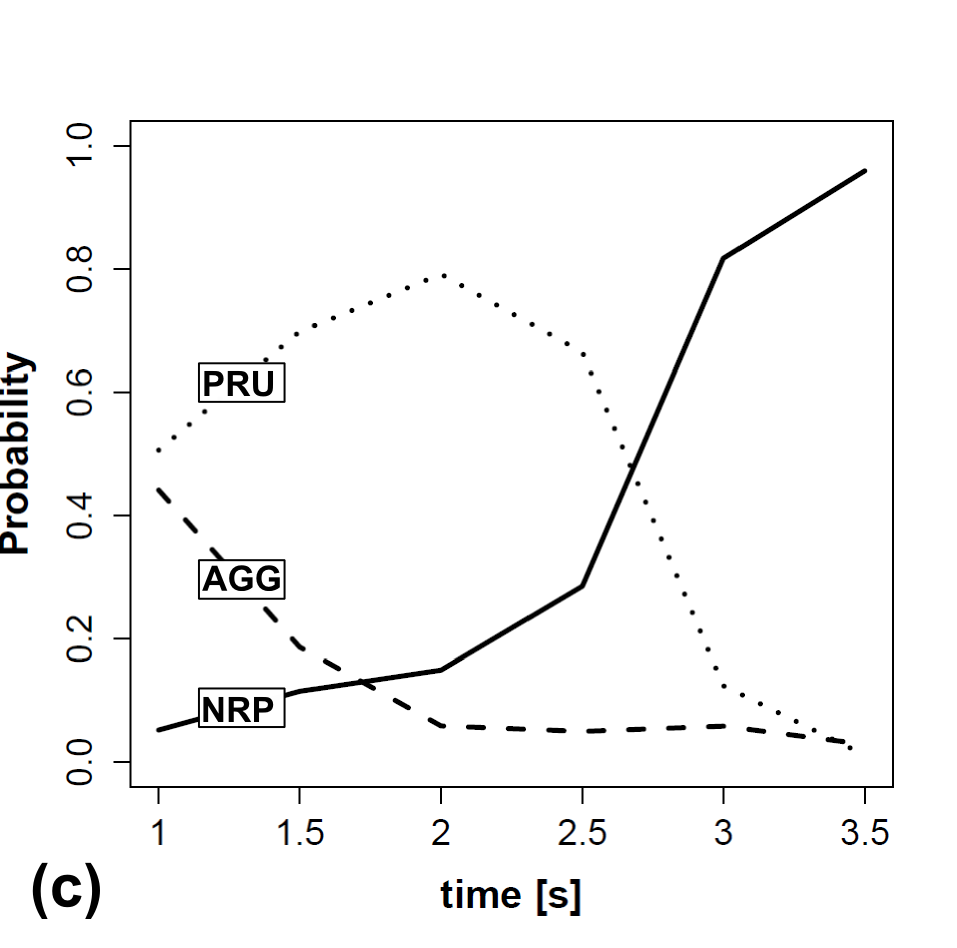}
	\caption{Situation 3: Characteristic values of pedestrian P3. The model expects P3 to be prudent instead of aggressive.}
	\label{fig:Sit3}
\end{figure}

However, the misclassification rate shows that in approximately 1 CI over 4 the model diverges from reality. For this reason - and in view of possible model improvements - many situations were tested and the reason of model misclassification was annotated. Two main causes were found. The first one is related to courtesy behavior - e.g. when a driver decelerates to let a pedestrian cross. In this case the model would classify the reaction as ACC or NRV, while the drivers actually decelerates (second column of Tab.\ref{tab:ConfusionMatrix} for vehicles). The second one is closely related to the heterogeneity of pedestrian behavior. This is evident in Tab.\ref{tab:ConfusionMatrix} for pedestrians where the AGG row and column have both high number of elements. One can think of elderly people which prefer to give priority to vehicles even if they could cross safely (elderly people have lower levels of risk acceptance). On the contrary, young people tend to be less prudent and to accept higher risks. This very case is shown in Fig.\ref{fig:Sit3}, where the upcoming car $v3$ (which is quite close and fast) is not captured by the video frame. While the observed behavior is classified as AGG for the first part of the reaction, the model expects pedestrian $P3$ to be prudent, meaning to let the vehicle pass.


\section{Conclusions and future research}
This paper investigated the behavior of road users in vehicle-pedestrian conflicts in shared space. Reaction mechanisms have been analyzed through real world observations and classified, and a statistical method has been developed to automatically detect the chosen reaction in video data at every time step. Successively, geometrical variables which affect the choice of the evasive action have been investigated and included in the multinomial logit model. The model has been calibrated through a large dataset of observed data and has shown good performances given the simple mathematical structure and the restricted number of variable.  
However, two main aspects must be considered for model improvement. First, the relevance of age and time pressure in pedestrian behavior. Second, the high frequency of courtesy behavior in shared space, i.e. a vehicle letting a pedestrian cross while standing at the side of the road. Moreover, considering only a single scenario, the statistical analysis does not ensure spacial transferability of the model. The calibration of the model in other shared space streets can reveal in fact the influence of the road design and regulation in pedestrian and vehicle conflict behavior. Finally, the model can be adapted to other traffic scenarios like mid-block crossings or refuge islands, where priority rules are clearly defined but interaction always occur. 
With the purpose of microsimulation, the developed model is ready-to-use and can be directly implemented in any motion model. This is actually part of future research within the project MODIS, where the aim will be to extend the existing model for capturing conflict situations involving many road users.

\section{Acknowledgement}
The scientific research published in this article is granted by the DFG under the reference BE 2159/13-1 and FR 1670/13-1. The authors cordially thank the funding agency.


\bibliography{library}

\begin{thebibliography}{10}

\bibitem{Anvari2014}
B.~Anvari, M.~G. Bell, P.~Angeloudis, and W.~Y. Ochieng.
\newblock Long-range collision avoidance for shared space simulation based on
  social forces.
\newblock {\em Transportation Research Procedia}, 2:318--326, 2014.

\bibitem{Anvari2015}
B.~Anvari, M.~G. Bell, A.~Sivakumar, and W.~Y. Ochieng.
\newblock {Modelling shared space users via rule-based social force model}.
\newblock {\em Transportation Research Part C: Emerging Technologies},
  51:83--103, 2015.

\bibitem{FGSV2014}
R.~Baier.
\newblock {Hinweise zu Stra{\ss}enr\"aumen mit besonderem Querungsbedarf -
  Anwendungsm\"oglichkeiten des "Shared Space" - Gedankens}.
\newblock Technical report, Forschungsgesellschaft f\"ur Stra{\ss}en- und
  Verkehrswesen, 2014.

\bibitem{DepartmentofTransport2011}
{Department of Transport}.
\newblock {Shared Space}.
\newblock Technical report, {Local Transport Note 1/11. 2011}.

\bibitem{Helbing1995}
D.~Helbing and P.~Molnar.
\newblock {Social force model for pedestrian dynamics}.
\newblock {\em Physical Review E}, 51(5):4282, jan 1995.

\bibitem{Himanen1988}
V.~Himanen and R.~Kulmala.
\newblock {An application of logit models in analysing the behaviour of
  pedestrians and car drivers on pedestrian crossings}.
\newblock {\em Accident Analysis {\&} Prevention}, 20(3):187--197, 1988.

\bibitem{Flow2012}
M.~Joyce.
\newblock {Shared Space in Urban Environments}.
\newblock Technical report, Flow Transportation Specialists. Guidance Note
  7/11, 2011.

\bibitem{Kaparias2016}
I.~Kaparias, J.~Hirani, M.~G. Bell, and B.~Mount.
\newblock Pedestrian gap acceptance behavior in street designs with elements of
  shared space.
\newblock {\em Transportation Research Record: Journal of the Transportation
  Research Board}, 2586:17--27, 2016.

\bibitem{Oxley2005}
J.~A. Oxley, E.~Ihsen, B.~N. Fildes, J.~L. Charlton, and R.~H. Day.
\newblock {Crossing roads safely: An experimental study of age differences in
  gap selection by pedestrians}.
\newblock {\em Accident Analysis {\&} Prevention}, 37(5):962--971, 2005.

\bibitem{Pascucci2015}
F.~Pascucci, N.~Rinke, C.~Schiermeyer, B.~Friedrich, and V.~Berkhahn.
\newblock {Modeling of shared space with multi-modal traffic using a
  multi-layer social force approach}.
\newblock {\em Transportation Research Procedia}, 10:316--326, 2015.

\bibitem{Rinke2016}
N.~Rinke, C.~Schiermeyer, F.~Pascucci, V.~Berkhahn, and B.~Friedrich.
\newblock A multi-layer social force approach to model interactions in shared
  spaces using collision prediction.
\newblock {\em Transportation Research Procedia}, 25:1249--1267, 2017.

\bibitem{Schiermeyer2016}
C.~Schiermeyer, F.~Pascucci, N.~Rinke, V.~Berkhahn, and B.~Friedrich.
\newblock A genetic algorithm approach for the calibration of a social force
  based model for shared spaces.
\newblock {\em Proceedings of the 8th International Conference on Pedestrian
  and Evacuation Dynamics (PED)}, 2016.

\bibitem{Schonauer2012}
R.~Sch{\"{o}}nauer, M.~Stubenschrott, W.~Huang, C.~Rudloff, and M.~Fellendorf.
\newblock {Modeling concepts for mixed traffic: Steps towards a microscopic
  simulation tool for shared space zones}.
\newblock {\em Transportation Research Record: Journal of the Transportation
  Research Board}, 43:1--16, 2012.

\bibitem{Schroeder2008}
B.~J. Schroeder.
\newblock {\em A behavior-based methodology for evaluating pedestrian-vehicle
  interaction at crosswalks}.
\newblock North Carolina State University, 2008.

\bibitem{Sun2002}
D.~Sun.
\newblock Modeling of motorist-pedestrian interaction at uncontrolled mid-block
  crosswalks. presented at 82rd annual meeting of the transportation research
  board, washington, d.c.
\newblock 2003.

\bibitem{Varhelyi1998}
A.~V{\'{a}}rhelyi.
\newblock {Drivers' speed behaviour at a zebra crossing: A case study}.
\newblock {\em Accident Analysis and Prevention}, 30(6):731--743, 1998.

\end{thebibliography}

\bibliographystyle{abbrv}

\end{document}